\documentclass[aps,amssymb,amsmathprx,reprint,preprintnumbers,superscriptaddress,nofootinbib,longbibliography,floatfix]{revtex4-2}

\usepackage{graphicx}
\usepackage{dcolumn}
\usepackage{bm}
\usepackage{hyperref}
\usepackage[caption=false]{subfig}
\hypersetup{
  colorlinks=true,
  citecolor=blue,
  linkcolor=blue,
  urlcolor=blue
}

\DeclareRobustCommand{\Sec}[1]{Sec.~\ref{sec:#1}}
\DeclareRobustCommand{\Secs}[2]{Secs.~\ref{sec:#1} and \ref{sec:#2}}
\DeclareRobustCommand{\App}[1]{App.~\ref{app:#1}}

\DeclareRobustCommand{\Fig}[1]{Fig.~\ref{fig:#1}}

\DeclareRobustCommand{\Eq}[1]{Eq.~(\ref{eq:#1})}

\DeclareRobustCommand{\Reference}[1]{Ref.~\cite{#1}}
\DeclareRobustCommand{\Refs}[1]{Refs.~\cite{#1}}

\usepackage{color}
\definecolor{darkred}{rgb}{1.0,0.1,0.1}
\definecolor{darkgreen}{rgb}{0.1,0.7,0.1}
\definecolor{darkblue}{rgb}{0.1,0.1,1.0}

\begin{document}

\preprint{MIT-CTP/5641}

\title{Safe but Incalculable: Energy-weighting is not all you need}

\author{Samuel Bright-Thonney}
\email{skb93@cornell.edu}
\affiliation{Physics Department, Cornell University, 109 Clark Hall, Ithaca, New York 14853, USA}

\author{Benjamin Nachman}
\email{bpnachman@lbl.gov}
\affiliation{Physics Division, Lawrence Berkeley National Laboratory, Berkeley, CA 94720, USA}
\affiliation{Berkeley Institute for Data Science, University of California, Berkeley, CA 94720, USA}

\author{Jesse Thaler}
\email{jthaler@mit.edu}
\affiliation{Center for Theoretical Physics, Massachusetts Institute of Technology, Cambridge, MA 02139, USA}
\affiliation{The NSF AI Institute for Artificial Intelligence and Fundamental Interactions, U.S.A.}

\begin{abstract}
Infrared and collinear (IRC) safety has long been used a proxy for robustness when developing new jet substructure observables.
This guiding philosophy has been carried into the deep learning era, where IRC-safe neural networks have been used for many jet studies.
For graph-based neural networks, the most straightforward way to achieve IRC safety is to weight particle inputs by their energies.
However, energy-weighting by itself does not guarantee that perturbative calculations of machine-learned observables will enjoy small non-perturbative corrections.
In this paper, we demonstrate the sensitivity of IRC-safe networks to non-perturbative effects, by training an energy flow network (EFN) to maximize its sensitivity to hadronization.
We then show how to construct Lipschitz Energy Flow Networks ($L$-EFNs), which are both IRC safe and relatively insensitive to non-perturbative corrections.
We demonstrate the performance of $L$-EFNs on generated samples of quark and gluon jets, and showcase fascinating differences between the learned latent representations of EFNs and $L$-EFNs.
\end{abstract}

\maketitle

\tableofcontents

\section{\label{sec:intro}Introduction}

Infrared and collinear (IRC) safety has played a central role in perturbative quantum chromodynamics (QCD)~\cite{Kinoshita:1962ur,Lee:1964is,Sterman:1977wj,Sterman:1978bi,Sterman:1978bj,CTEQ:1993hwr,Weinberg:1995mt,Banfi:2004yd,Sterman:2006uk,Komiske:2020qhg}, in particularly for studying the properties of high-energy jets arising from the fragmentation of quarks and gluons.
By being insensitive to soft and collinear splittings within a parton shower, an IRC-safe observable has a cross section that can be well-described by a fixed-order perturbation series in the strong coupling constant, $\alpha_s$.
For this reason, IRC safety has been used as a guiding principle in developing new jet substructure observables~\cite{Abdesselam:2010pt,Altheimer:2012mn,Altheimer:2013yza,Adams:2015hiv,Larkoski:2017jix,Kogler:2018hem,Marzani:2019hun,Kogler:2021kkw}.
While other classes of calculable observables exist---such as Sudakov-safe observables~\cite{Larkoski:2013paa,Larkoski:2015lea} and track-based observables~\cite{Chang:2013rca,Li:2021zcf}\footnote{More generally, there is a broader class of observables that can be calculated after absorbing perturbative singularities into universal non-perturbative objects like fragmentation functions~\cite{Altarelli:1981ax,Elder:2017bkd}.}---IRC safety has driven the selection of collider observables for both measurements and searches involving jets.

Machine learning (ML) methods have significantly extended the sensitivity of jet tagging, in part by leveraging low-level information~\cite{Larkoski:2017jix,Kasieczka:2019dbj,Feickert:2021ajf}.
Such taggers are generally IRC unsafe, which means that they are reliant on the non-perturbative models in the generators used to train and test them.
Since these non-perturbative models are imperfectly constrained by experimental data, this has motivated the development of ML observables that are IRC (or Sudakov) safe~\cite{Komiske:2017aww,Komiske:2018cqr,Athanasakos:2023fhq}.
One class of IRC-safe architectures are Energy Flow Networks (EFNs)~\cite{Komiske:2018cqr}, which are built on the deep sets framework~\cite{NIPS2017_f22e4747} and are able to model any permutation-invariant IRC-safe observable.
EFNs, along with extensions that incorporate graph-like structures~\cite{Konar:2021zdg,Shen:2023ofd,Konar:2023ptv}, achieve IRC safety by linearly weighting particle inputs by their energies.

In this paper, we emphasize that energy-weighting, by itself, does not ensure trustworthy perturbative calculations for ML observables.
While IRC safety does ensure the existence of a fixed-order perturbation series in $\alpha_s$,%
\footnote{Strictly speaking, this statement depends on the precise definition one uses for ``IRC safety''.  In the language of \Reference{Komiske:2020qhg}, we really mean H\"older continuous.}
it does not guarantee that non-perturbative corrections will be small.
Using EFNs as a representative example, we show how to train an IRC-safe neural network to be maximally sensitive to non-perturbative hadronization, thereby constructing an observable that is ``safe but incalculable''.\footnote{The opposite case of ``unsafe but calculable'' observables can arise in resummed perturbation theory, where there is no order-by-order $\alpha_s$ expansion but nevertheless non-perturbative corrections are suppressed~\cite{Larkoski:2013paa,Larkoski:2015lea}.}
As a step towards restoring calculablity, we introduce Lipschitz Energy Flow Networks ($L$-EFNs), whose bounded gradients ensure bounded sensitivity to non-perturbative corrections.

The fact that IRC-safe observables can have cross sections with large non-perturbative corrections is not new, even if it may not be widely appreciated.
The standard (but misleading) lore is that IRC-safe observables should have non-perturbative corrections that are power-suppressed as $(\Lambda_{\rm QCD}/E)^n$, where $\Lambda_{\rm QCD}$ is the QCD confinement scale, $E$ is the energy scale of the process in consideration, and $n$ is some integer power (typically 1 or 2).
Already, though, it is known that jet angularities~\cite{Ellis:2010rwa,Larkoski:2014uqa,Almeida:2008yp,Larkoski:2014pca} with angular exponent $\beta \lesssim 1$ have non-perturbative corrections with $n = \beta$ scaling~\cite{Manohar:1994kq,Banfi:2004yd,Larkoski:2013paa,Larkoski:2013paa}, which turns into $\mathcal{O}(1)$ effects as $\beta \to 0$.
Because there is no general first-principles understanding of non-perturbative QCD effects, then the cross section is essentially incalculable (or at least untrustable) if these corrections grow large.

In the context of IRC-safe ML models, we are not aware of any previous studies of the general impact of non-perturbative effects.
Here, to identify ML observables with maximal non-perturbative sensitivity, we train an IRC-safe classifier to distinguish parton-level from hadron-level events.
Classifiers whose cross sections have controlled non-perturbative corrections should be unable to distinguish between these samples.
Instead, we find that EFNs are highly effective at parton-level versus hadron-level classification, implying large non-perturbative sensitivity.
Our new $L$-EFN architecture reduces this sensitivity by imposing spectral normalization~\cite{miyato2018spectral,gouk2020regularisation}, which is equivalent to bounding the Lipschitz norm of the network (see related work in \Refs{anil2019sorting,Kitouni:2021fkh}).
This approach is motivated by the Kantorovich-Rubinstein duality theorem~\cite{KR:58} and the Energy Mover's Distance (EMD)~\cite{Komiske:2019fks}, which provides a robust way to estimate the size of non-perturbative effects.

The remainder of this paper is organized as follows.
In \Sec{method}, we introduce $L$-EFNs and explain how the Lipschitz constraint enforces an EMD bound on non-perturbative corrections.
We then perform a case study in \Sec{casestudy} to compare the hadronization sensitivity of EFNs and $L$-EFNs.
We investigate the learned latent representations of ($L$-)EFNs in \Sec{efn-filters} and conclude in \Sec{conclusions}.
For completeness, we perform a quark/gluon discrimination study in \App{quarkgluon}.

\section{Methodologies}
\label{sec:method}

\subsection{Lipschitz Energy Flow Networks}

The $L$-EFN architecture we propose in this work is built on top of a standard EFN, which provides a generic framework for learning IRC-safe observables.
Given a jet with constituent momenta $p_1,p_2,\ldots,p_M$, an EFN computes a function of the form:
\begin{equation}
\label{eq:EFN}
    \mathrm{EFN}(\{p_1,\ldots,p_M\}) = F\left(\sum_{i=1}^M z_i\Phi(\hat{p}_i)\right),
\end{equation}
where $z_i = p_{T,i}/p_{T,\mathrm{jet}}$ is the constituent momentum or energy fraction and $\hat{p}_i$ is the particle's angular position relative to the jet axis.
The function $\Phi : \mathbb{R}^2 \to \mathbb{R}^\ell$ maps individual particles to a latent space of dimension $\ell$.
The function $F : \mathbb{R}^\ell \to \mathbb{R}^{d_\mathrm{out}}$ maps the latent representation to the final output.
In a standard EFN, the functions $\Phi$ and $F$ are unconstrained and typically implemented as neural networks.
The additive and energy-weighted structure of an EFN guarantees a naturally permutation-invariant and IRC safe output; see \Reference{Komiske:2018cqr} for further discussion.

An $L$-EFN extends the EFN setup by constraining $\Phi$ and $F$ to be $L$-Lipschitz, meaning that
\begin{eqnarray*}
    \|\Phi(\hat{p}_1)-\Phi(\hat{p}_2) \| &\leq& L \| \hat{p}_1 - \hat{p}_2 \|, \\ 
    \| F(x_1) - F(x_2) \| &\leq& L \| x_1 - x_2 \|.
\end{eqnarray*}
This is effectively a bound on the gradients of these functions, though the Lipschitz constraint does not require $\Phi$ and $F$ to be everywhere differentiable.
In principle, one could choose different $L$ values for $\Phi$ and $F$, but we keep them the same for simplicity of discussion.

If $\Phi$ and $F$ are neural networks with $L$-Lipschitz activations,%
\footnote{Many standard activation functions are $1$-Lipschitz, such as ReLU, LeakyReLU, and Sigmoid.}
this amounts to a constraint on the \textit{spectral norm} of their weight matrices $W^i$ \cite{miyato2018spectral}:
\begin{equation}
    \sigma(W^i)  := \max_{\mathbf{h}\neq \mathbf{0}}\frac{\|W^i\mathbf{h}\|_2}{\|\mathbf{h}\|_2} \leq L.
\end{equation}
This can be enforced during training by scaling the weight matrices as $W^i \to LW^i/\sigma(W^i)$ using a computationally efficient estimation of $\sigma(W^i)$~\cite{miyato2018spectral}.
We focus on the $L = 1$ case throughout this paper, and $L$-EFN should be henceforth understood as $L=1$.

For the studies in \Sec{results} we use the base architectures from the \textsc{EnergyFlow} package \cite{Komiske:2018cqr,Komiske:2017aww}, implemented and trained using \textsc{TensorFlow}~\cite{tensorflow}, \textsc{Keras}~\cite{chollet2015keras}, and \textsc{Adam}~\cite{adam}.
To enforce the $1$-Lipschitz constraint when training $L$-EFNs, we replace all linear \texttt{Dense} layers in the networks with \texttt{SpectralDense} layers from the \textsc{deel-lip} package \cite{2006.06520}.
Unless otherwise specified, the $\Phi$ ($F$) networks of all ($L$-)EFNs are implemented as three-layer fully-connected neural networks with a width of 60 (80) and 0\% (10\%) dropout~\cite{JMLR:v15:srivastava14a}.
We use ReLU activations for internal nodes and Sigmoid for the 1D output node, both of which have a Lipschitz constant of 1.
The networks are trained to learn 0/1 class labels based on the binary cross-entropy loss function.

As a reference, we also show results for a particle flow network (PFN), which is IRC unsafe by construction:
\begin{equation}
    \mathrm{PFN}(\{p_1,\ldots,p_M\}) = F\left(\sum_{i=1}^M \Phi(p_i)\right).
\end{equation}
This differs from the EFN in \Eq{EFN} by the lack of $z_i$ weighting and the use of the unnormalized particle momentum $p_i$ instead of just the particle direction $\hat{p}_i$.
We also consider a Lipschitz PFN ($L$-PFN) which imposes a spectral norm constraint on $F$ and $\Phi$.

\subsection{Bounds on the Energy Mover's Distance}

With the simple addition of a Lipschitz constraint, we can now make precise statements about the sensitivity of an $L$-EFN to non-perturbative effects.
Here, we leverage a recent geometric language for analyzing collision events based on the EMD~\cite{Komiske:2019fks}, which measures the ``work" required to transform one jet into another.
The EMD is a variant of the Earth Mover's Distance from computer vision~\cite{DBLP:journals/pami/PelegWR89,Rubner:1998:MDA:938978.939133,Rubner:2000:EMD:365875.365881,DBLP:conf/eccv/PeleW08,DBLP:conf/gsi/PeleT13} and is equivalent to the 1-Wasserstein metric~\cite{wasserstein1969markov,dobrushin1970prescribing} in certain limits.
The EMD provides a pairwise metric distance for IRC-safe energy flows, which can be used to triangulate the space of jets and define various geometric quantities.

Consider an additive IRC-safe observable $f$ acting on a jet $\mathcal{J}$:
\begin{equation}
f(\mathcal{J}) = \sum_{i \in \mathcal{J}}E_i\Phi(\hat{p}_i),
\end{equation}
where $\Phi$ has Lipschitz constant $L$.
Given a pair of jets $\mathcal{J}_1$, $\mathcal{J}_2$, a key result from \Reference{Komiske:2019fks} places a bound on the difference between $f(\mathcal{J}_1)$ and $f(\mathcal{J}_2)$ based on the EMD between $\mathcal{J}_1$ and $\mathcal{J}_2$:
\begin{equation}
    \frac{1}{RL}\left| f(\mathcal{J}_1) - f(\mathcal{J}_2)\right| \leq \mathrm{EMD}(\mathcal{J}_1,\mathcal{J}_2),
\end{equation}
where $R$ is the jet radius.
The composition of two Lipschitz functions with constants $L_A$ and $L_B$ is also Lipschitz, with constant $L_A L_B$.
Thus, for an $L$-EFN, we have:
\begin{equation}
    \label{eq:efsn-bound}
    \frac{1}{RL^2} \left| L\text{-EFN}(\mathcal{J}_1) - L\text{-EFN}(\mathcal{J}_2)\right| \leq \mathrm{EMD}(\mathcal{J}_1,\mathcal{J}_2).
\end{equation}
This bound holds for any $\mathcal{J}_1$ and $\mathcal{J}_2$, and we again emphasize that we focus on $L=1$ in this paper.

Now, consider $\mathcal{J}_1$ and $\mathcal{J}_2$ to be the \textit{same jet} at parton- and hadron-level, respectively.
Of course, parton-level information cannot be accessed at a real collider, but it can be done in the context of a parton shower event generator.
Then, \Eq{efsn-bound} gives an upper bound on modifications of the $L$-EFN observable due to hadronization.
This prevents an $L$-EFN from becoming overly sensitive to non-perturbative physics, and makes it naturally robust to any unphysical mismodeling artifacts present in Monte Carlo simulations.
By contrast, a plain EFN does not have this constraint, which explain why it can be arbitrarily sensitive to non-perturbative effects, as we will see in the following case study.

\section{Case Study with Quark/Gluon Jets}
\label{sec:casestudy}

\subsection{Generated Dataset}
\label{sec:dataset}

To demonstrate the impact of non-perturbative physics on ML algorithms, we perform a case study involving quark and gluon jets.
Following \Reference{Bright-Thonney:2022xkx}, we generate samples of $e^+e^- \to H \to q\bar{q}$ (quark jets) and $gg$ (gluon jets) using \textsc{Pythia 8.307}.
We take a center-of-mass collision energy of $\sqrt{s} = m_H = 1$ TeV, and we fix $q=u$ (i.e.~up quarks only) and $m_u = 0$ for simplicity.

In each event, we independently cluster the partons (post-shower) and final-state hadrons into anti-$k_T$ jets with $R = 1.0$~\cite{Cacciari:2011ma,Cacciari:2008gp} and the winner-take-all (WTA) axis.
We match the leading (highest $p_T$) hadron-level jet to the nearest parton-level jet within $\Delta R < 0.1$,\footnote{If no parton-level jet is found within this $\Delta R$ cone, the event is skipped.} and save the constituent kinematic information for both jets.
We generate a total of 200,000 $e^+e \to q\bar{q}$ and 200,000 $e^+e^- \to gg$ events.
Each dataset is split into training, validation, and test sets with fractions 50\%, 25\%, and 25\%, respectively.

\begin{figure*}[t]
    \centering
    \subfloat[][]{
    \includegraphics[width=0.32\textwidth]{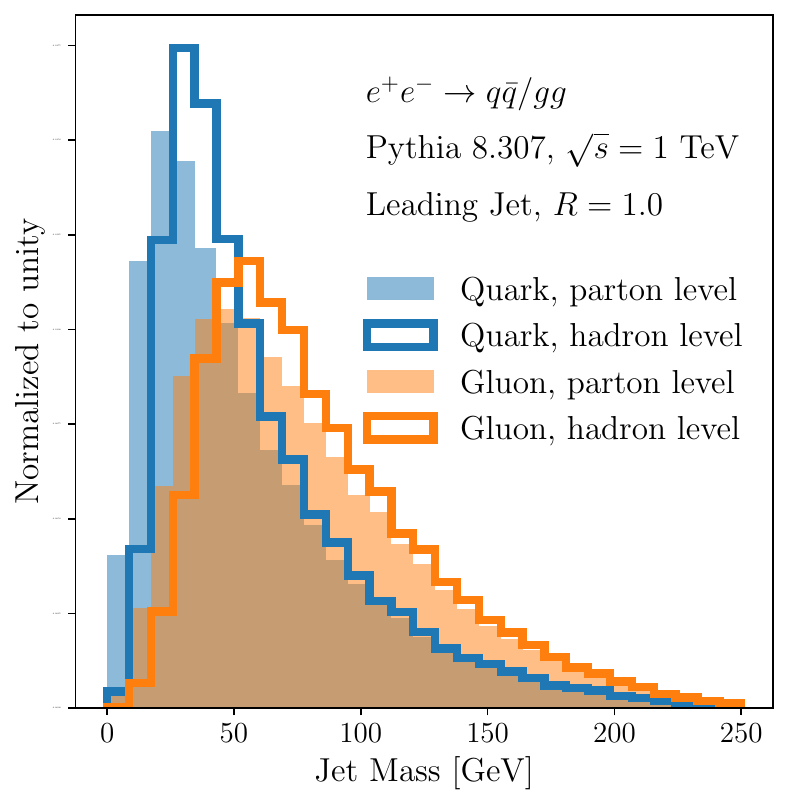}}
    \subfloat[][]{
    \includegraphics[width=0.32\textwidth]{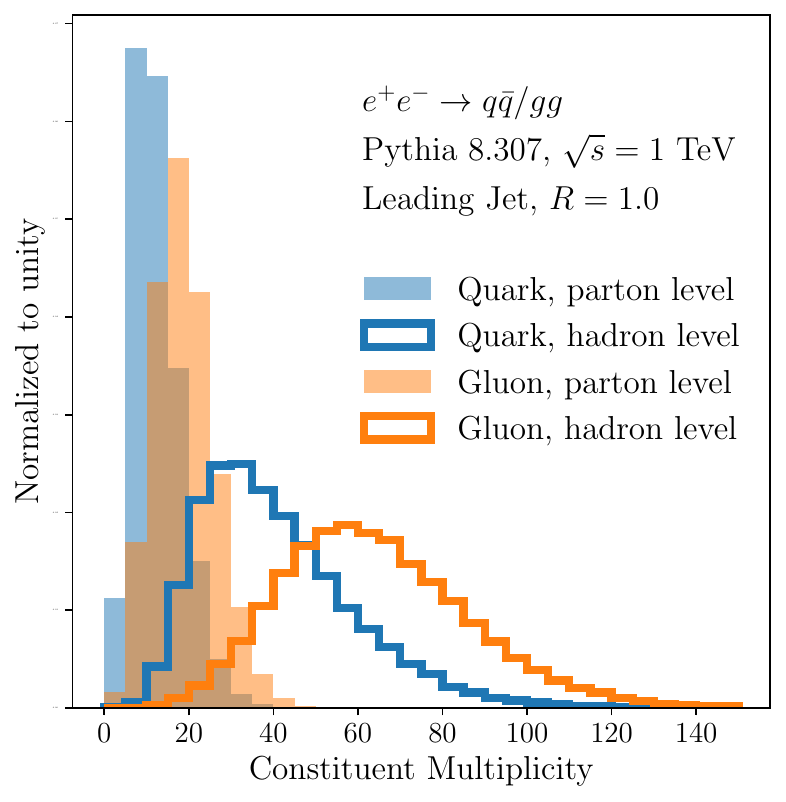}}
    \subfloat[][]{
    \includegraphics[width=0.32\textwidth]{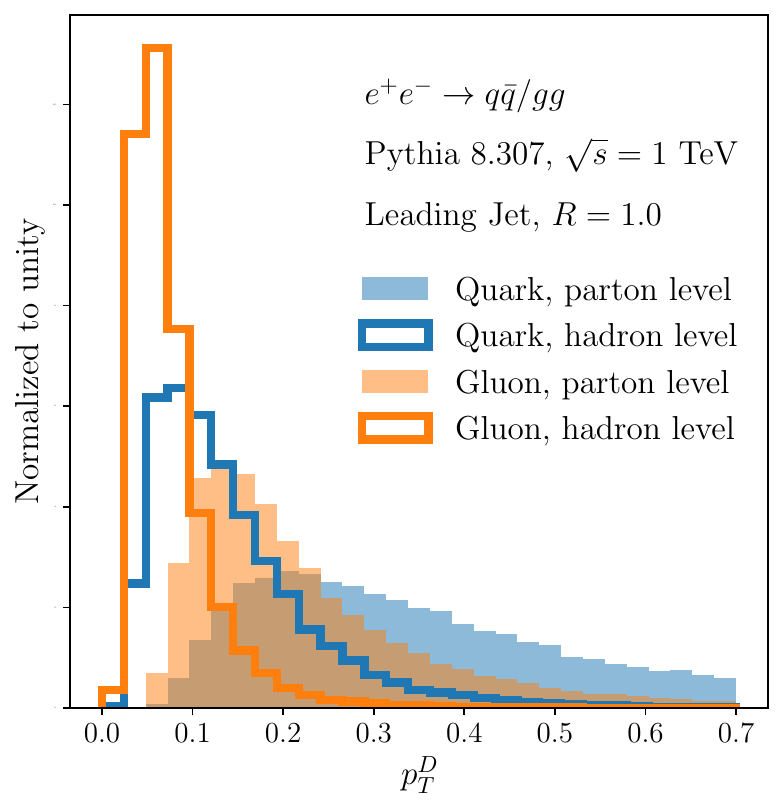}}
    \caption{
    The spectrum of (a) jet mass, (b) jet constituent multiplicity, and (c) $p_T^D$.  Shown are the normalized distributions for quarks (blue) and gluons (orange) at parton-level (filled) versus hadron-level (solid).
    }
    \label{fig:data}
\end{figure*}

\begin{figure*}[t]
    \centering
        \subfloat[][]{
    \includegraphics[width=0.38\textwidth]{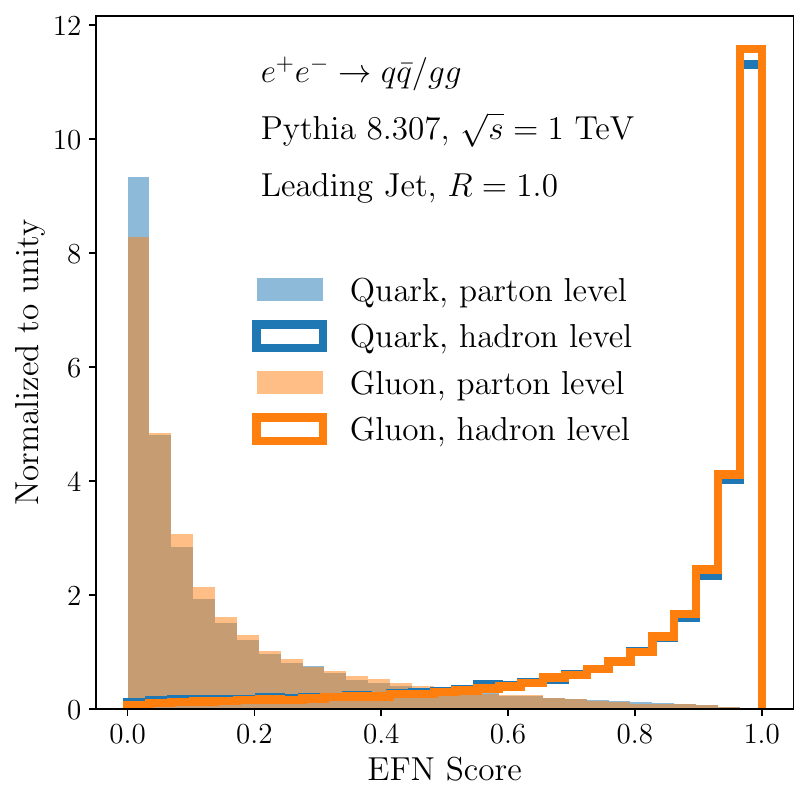}
    } $\qquad\qquad$
        \subfloat[][]{
    \includegraphics[width=0.38\textwidth]{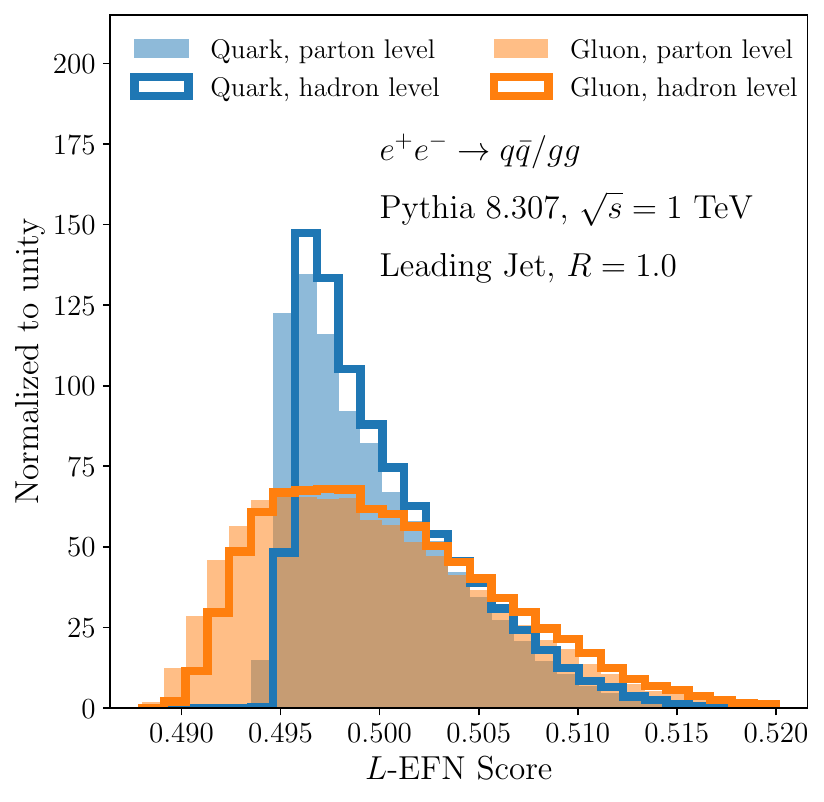}
    }
    \caption{Score distributions for the (a) EFN and (b) $L$-EFN architectures, when trained to distinguish parton-level from hadron-level jets.
    The color scheme is the same as \Fig{data}.
    Despite being IRC safe, the EFN achieves excellent parton/hadron separation power.
    After imposing the Lipschitz constraint, the $L$-EFN exhibits the desired insensitivity to non-perturbative effects.
    }
    \label{fig:EFN:dist}
\end{figure*}

\begin{figure*}[t]
    \centering
            \subfloat[][]{
    \includegraphics[width=0.4\textwidth]{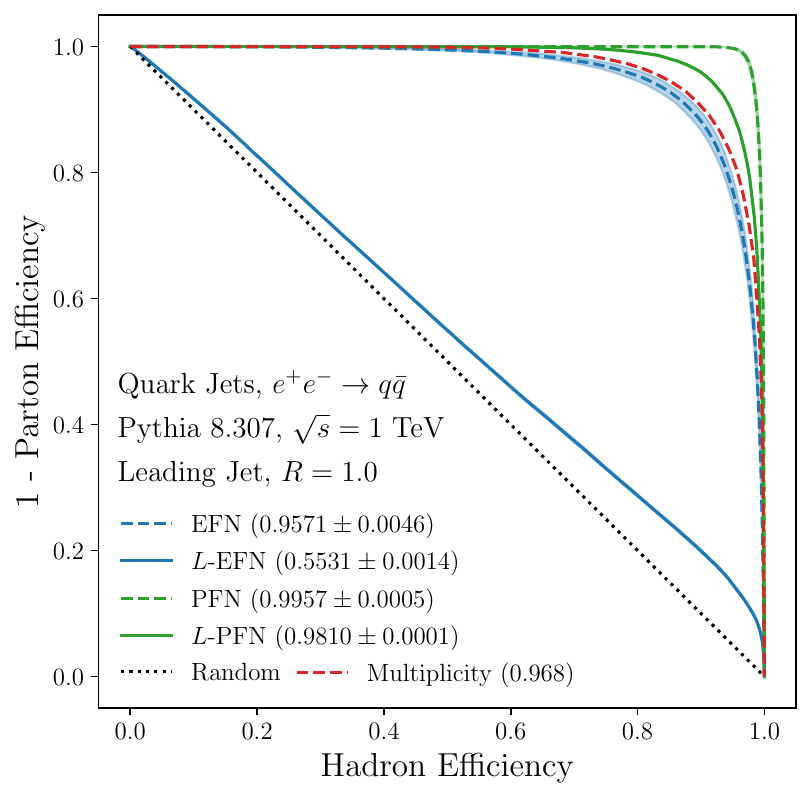}
    } $\qquad\qquad$
            \subfloat[][]{
    \includegraphics[width=0.4\textwidth]{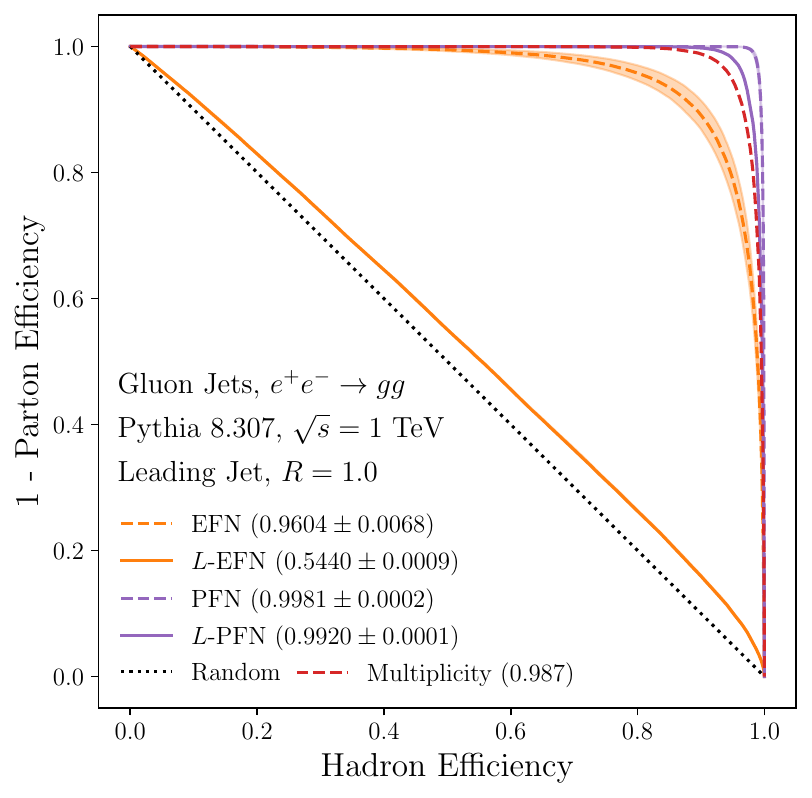}
    }
    \caption{
    ROC curves for parton- vs hadron-level discrimination for (a) quark and (b) gluon jets using an EFN/PFN (solid) and an $L$-EFN/$L$-PFN (dashed).
    The bands represents the spread from starting with 10 different random initializations of the neural networks.
    The results for constituent multiplicity is shown for reference.
    Only the $L$-EFN shows the desired lack of sensitivity to non-perturbative effects.
    }
    \label{fig:EFN:roc}
\end{figure*}

In \Fig{data}, we show the spectrum of three jet substructure observables: the jet mass, the number of jet constituents, and $p_T^D$~\cite{CMS:2012rth}.
The latter observable is the sum of the squares of the momentum fractions of constituents within the jet.
Of these observables, only the jet mass is IRC safe.
Not surprisingly, the jet mass is the observable with the smallest difference between the parton-level and hadron-level distributions.
For the number of constituents and $p_T^D$, non-perturbative corrections are an $\mathcal{O}(1)$ effect.
Nevertheless, these two observables are highly effective for quark versus gluon jet tagging~\cite{Gras:2017jty}, which is why one has to be careful about non-perturbative sensitivity in the ML context.

\subsection{Classification Results}
\label{sec:results}

To investigate the impact of Lipschitz constraints on non-perturbative sensitivity, we train EFNs and $L$-EFNs to distinguish hadron-level from parton-level jets.
For this analysis, we treat the quark and gluon datasets separately.
(See \App{quarkgluon} for a quark/gluon discrimination study.)
To verify the stability of our trainings and estimate uncertainties on performance metrics, we train ten versions of each model with a different random initializations.
In testing, we found that $L$-EFNs train more stably with a significantly larger batch size than EFNs (10,000 vs.\ 128) and a smaller learning rate ($10^{-4}$ vs.\ $10^{-3}$).
This is likely due to the spectral normalization of the weight matrices at each step, and the inherent difficulty of distinguishing parton-level from hadron-level jets given the constraint of \Eq{efsn-bound}. With these modifications, however, all $L$-EFN trainings converged easily.

Distributions of the learned EFN and $L$-EFN scores are shown in \Fig{EFN:dist}, considering all four combinations of quark versus gluon and parton-level versus hadron-level jets.
Despite being IRC safe, the EFN is highly effective at distinguishing parton-level from hadron-level jets.
This implies that the non-perturbative corrections to the EFN observable are large.
Assuming that the EFN architecture and training procedure are sufficiently flexible,\footnote{We did not perform an extensive hyperparameter scan, but we found little gains from small variations on our setup.} then this is an example of an IRC-safe observable with maximal non-perturbative sensitivity.
By contrast, the $L$-EFN is not very effective at distinguishing parton-level from hadron-level jet, so according to the hadronization model in \textsc{Pythia}, the non-perturbative corrections are small.

To better assess the learned separation power, we plot receiver operating characteristic (ROC) curves in \Fig{EFN:roc}, obtained by placing a sliding cut over the distributions in \Fig{EFN:dist}.
We also indicate their Area Under the Curve (AUC) scores in the legend.
The ROC curves underscore and quantify the trends observed in \Fig{EFN:dist}.
The EFN is an extremely efficient parton-versus-hadron classifier, whereas the $L$-EFN performs barely better than random chance.
We therefore conclude that the $1$-Lipschitz modification is effective at suppressing the $L$-EFN's sensitivity to non-perturbative effects.

As a point of comparison, \Fig{EFN:roc} also shows ROCs and AUCs for multiplicity, PFNs and $L$-PFNs -- all explicitly IRC-\textit{unsafe} observables.
Unsurprisingly, the PFN is able to nearly perfectly distinguish parton from hadron jets, and the $L$-PFN performs only slightly worse.
This indicates that the $1$-Lipschitz constraint does not severely limit the expressiveness of the model, and that the large gap between the EFN and $L$-EFN is a genuine result of non-perturbative effects.

\begin{figure*}[t]
    \centering
    \subfloat[][]{
    \includegraphics[width=0.32\textwidth]{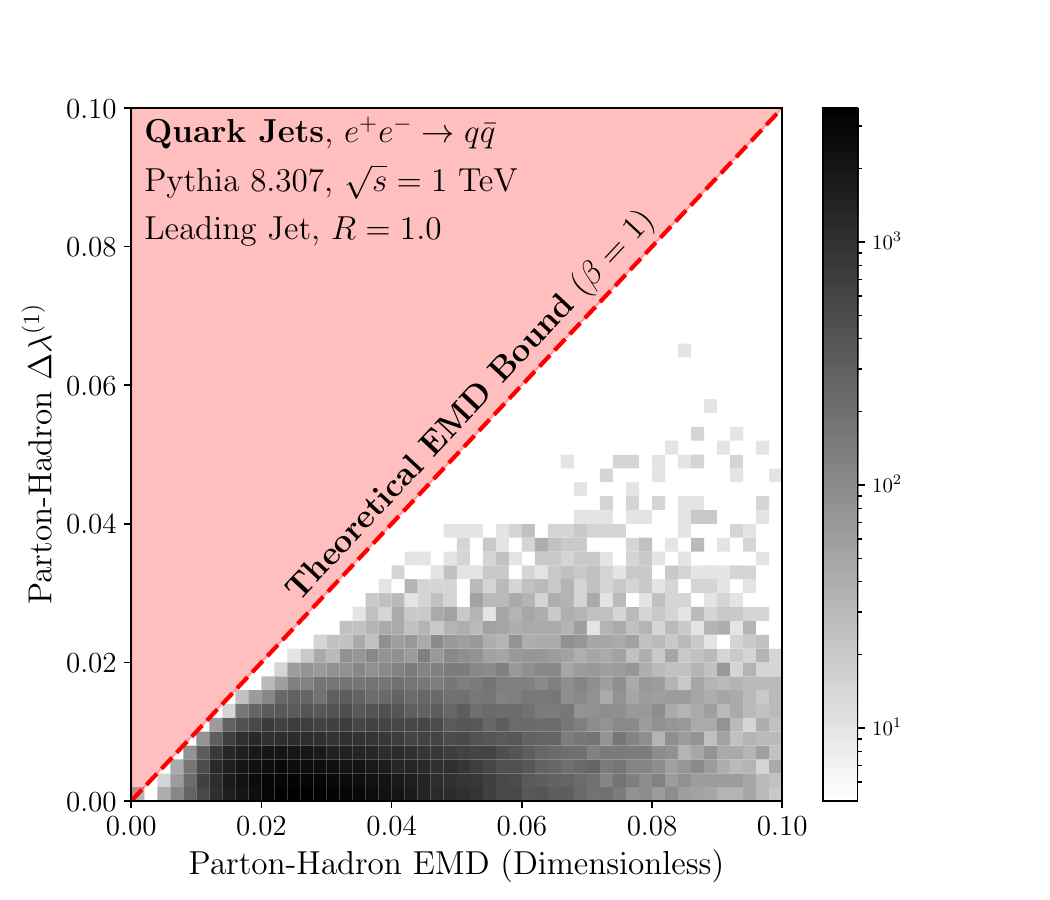}
    }
    \subfloat[][]{
    \includegraphics[width=0.32\textwidth]{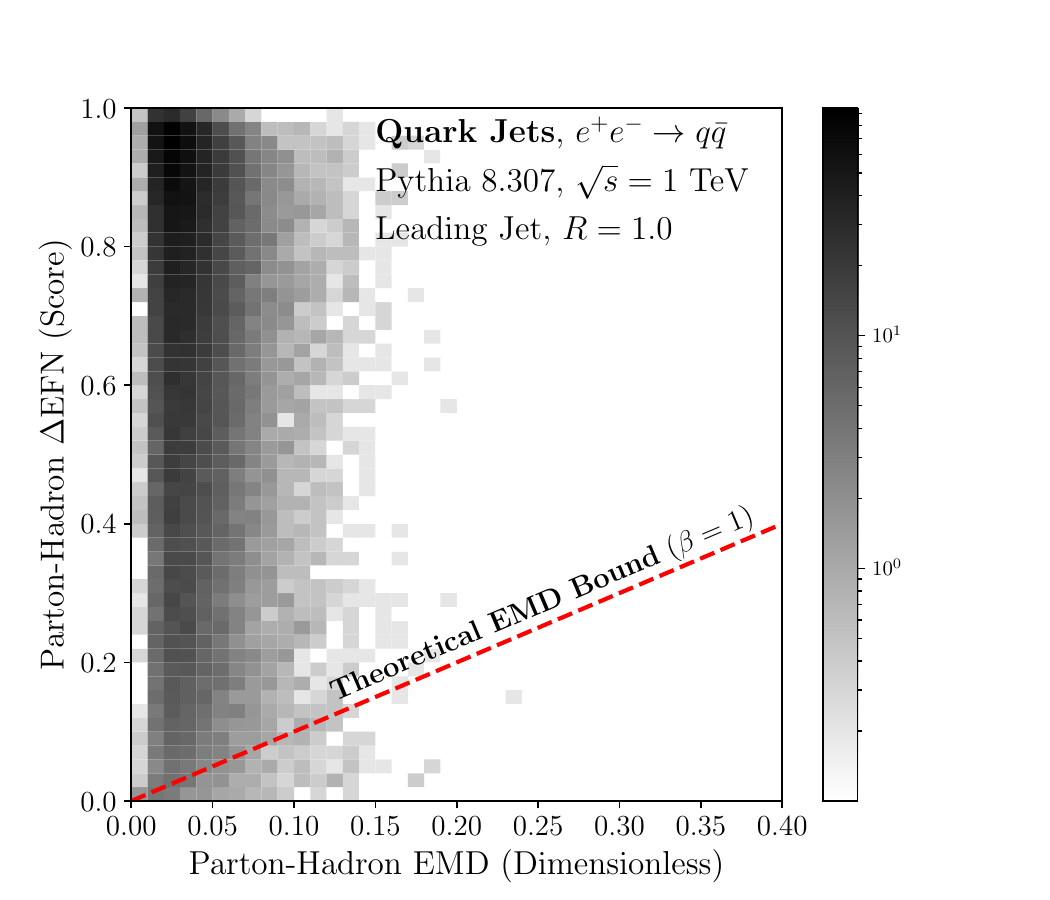}
    }
    \subfloat[][]{
    \includegraphics[width=0.32\textwidth]{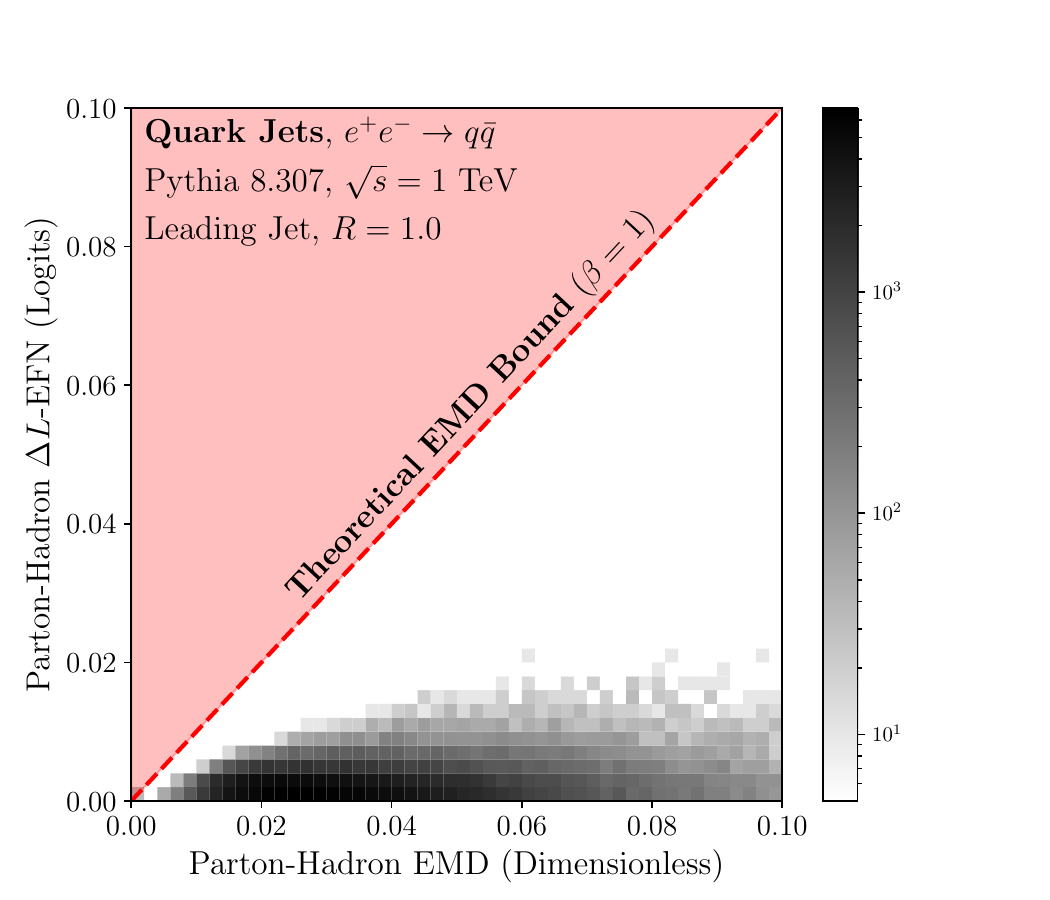}
    }
    \\
    \subfloat[][]{
    \includegraphics[width=0.32\textwidth]{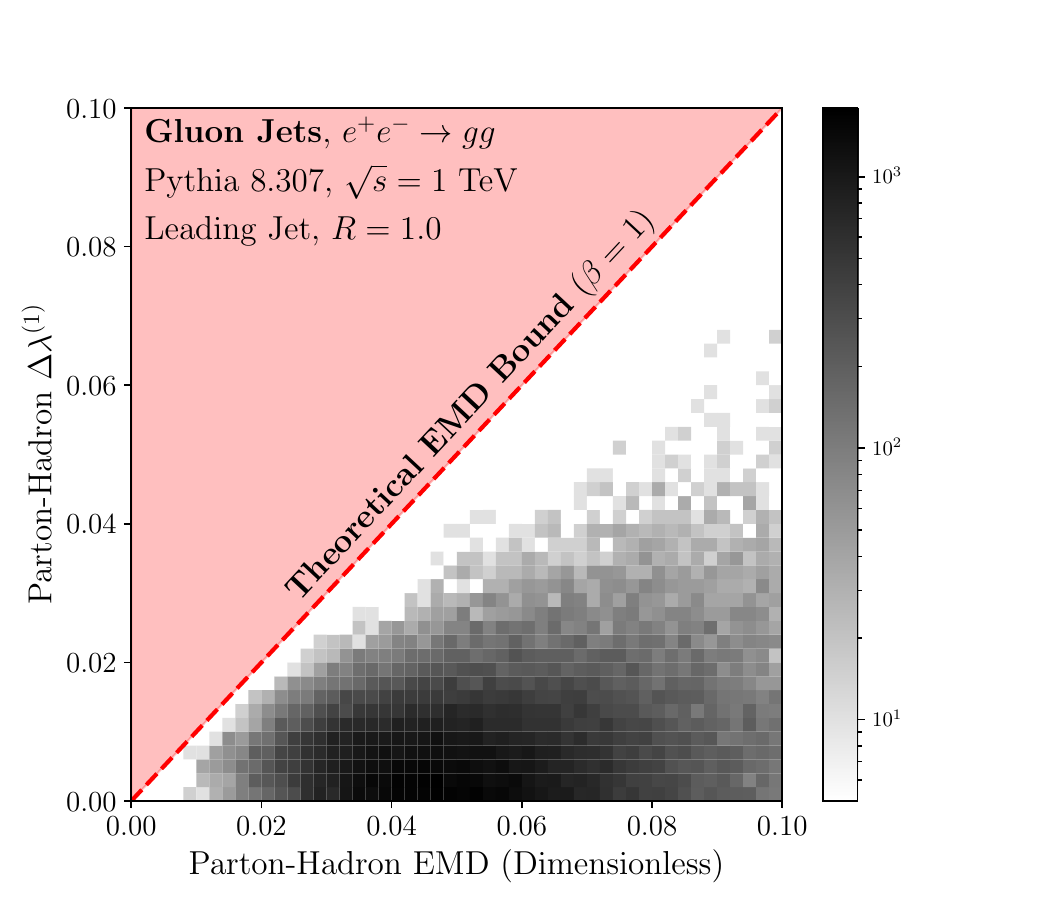}
    }
    \subfloat[][]{
    \includegraphics[width=0.32\textwidth]{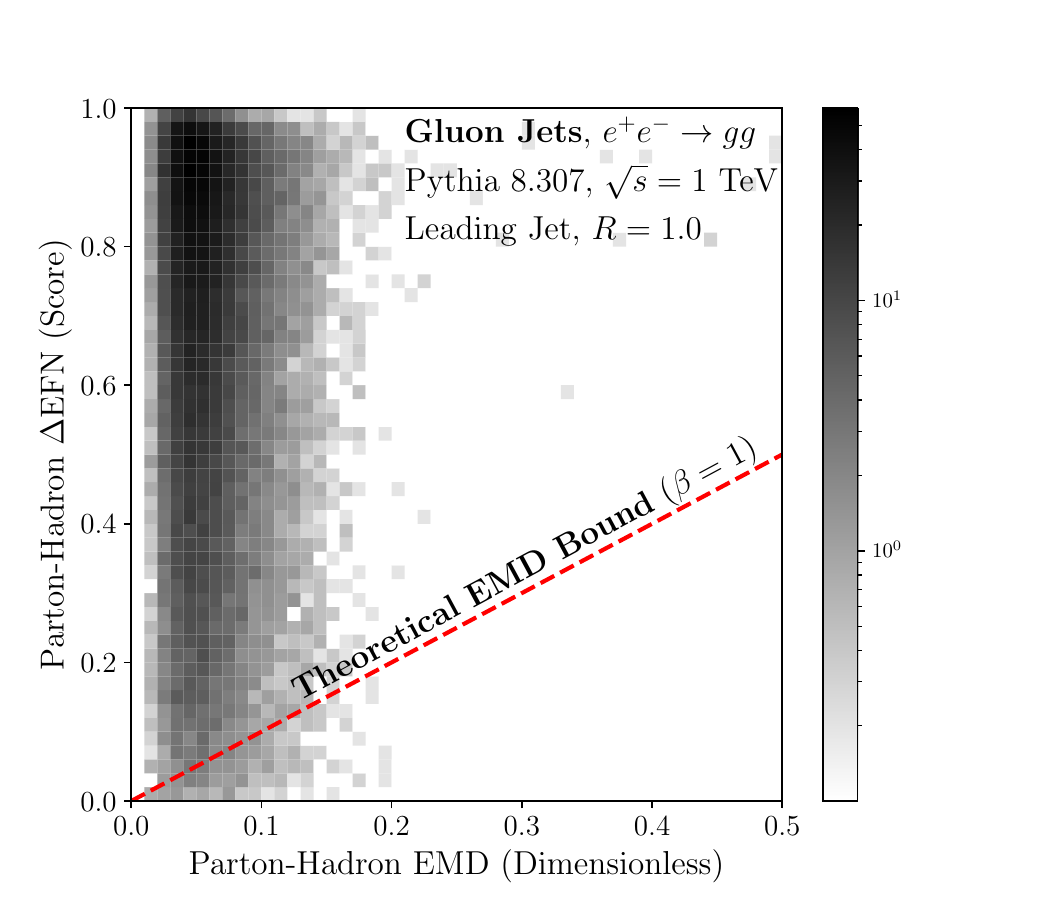}
    }
    \subfloat[][]{\includegraphics[width=0.32\textwidth]{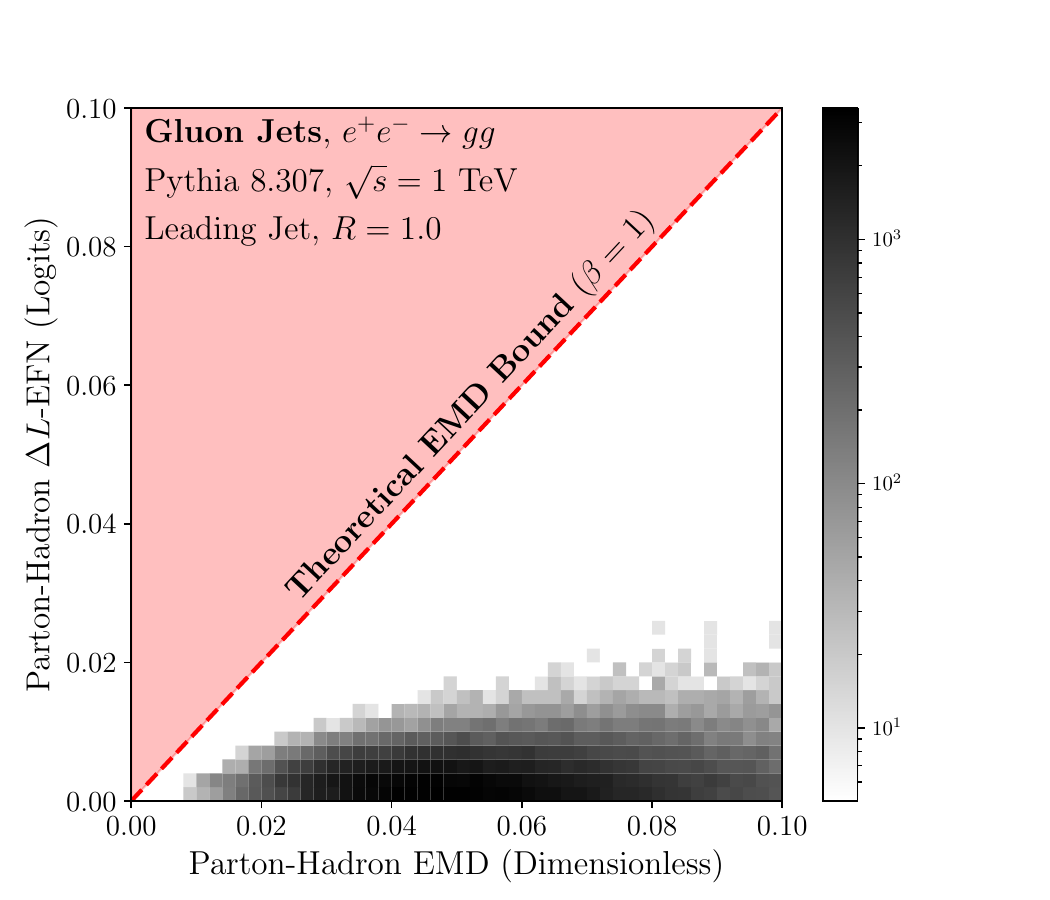}
    }
    \caption{Two dimensional distributions of the parton-hadron EMD versus the difference in
observables at parton- and hadron-level.
Shown are results for (left column) the $\beta = 1$ angularity $\lambda^{(1)}$, (middle column) an EFN, (right column) an $L$-EFN, separately for (top row) quark jets and (bottom row) gluon jets.
The EMD is computed using momentum fractions $z_i$ in keeping with the $\lambda^{(1)}$/EFN/$L$-EFN calculations, and is thus dimensionless.}
    \label{fig:emdplots}
\end{figure*}

\subsection{Confirming the EMD Bound}

As a cross-check of our analysis, we verify that the learned $L$-EFN satisfies the EMD bound in \Eq{efsn-bound}.
The plots in \Fig{emdplots} compares the difference between observables at parton level versus hadron level to the EMD between parton- and hadron-level events.

The top row of \Fig{emdplots} reproduces Fig.~2 of \Reference{Komiske:2019fks} for the $\beta = 1$ angularity:
\begin{equation}
\lambda^{(1)} = \sum_{i\in\text{jet}}z_i\,\Delta R_{i,\text{jet}},
\end{equation}
where $z_i = p_{T,i}/p_{T,\text{jet}}$ and $\Delta R_{i,\text{jet}}^2 = \Delta \eta_{i,\text{jet}}^2 + \Delta \phi_{i,\text{jet}}^2$.
This function is $1$-Lipschitz and the bound is respected and nearly saturated.

The middle row of \Fig{emdplots} shows the non-perturbative modification to the EFN score, which clearly violates the theoretical bound.
This is expected, since the EFN is not required to be a Lipschitz function.
Indeed, there is no finite slope (i.e.\ Lipschitz constant) that would bound the EFN score relative to the EMD.

By contrast, the bottom row of \Fig{emdplots} demonstrates that the $L$-EFN does respect the EMD bound.
For illustrative purposes, we show the $L$-EFN logits (i.e.\ pre-activations on the final layer), since the actual output scores all lie along the bottom of the plot.
Even after performing the logit transformation, the $L$-EFN does not appear to saturate the bound, unlike what was seen for the IRC-safe angularities.

\begin{figure*}[t]
    \centering
    \begin{minipage}[c]{0.45\textwidth}
        \subfloat[][]{
        \includegraphics[width=0.5\textwidth]{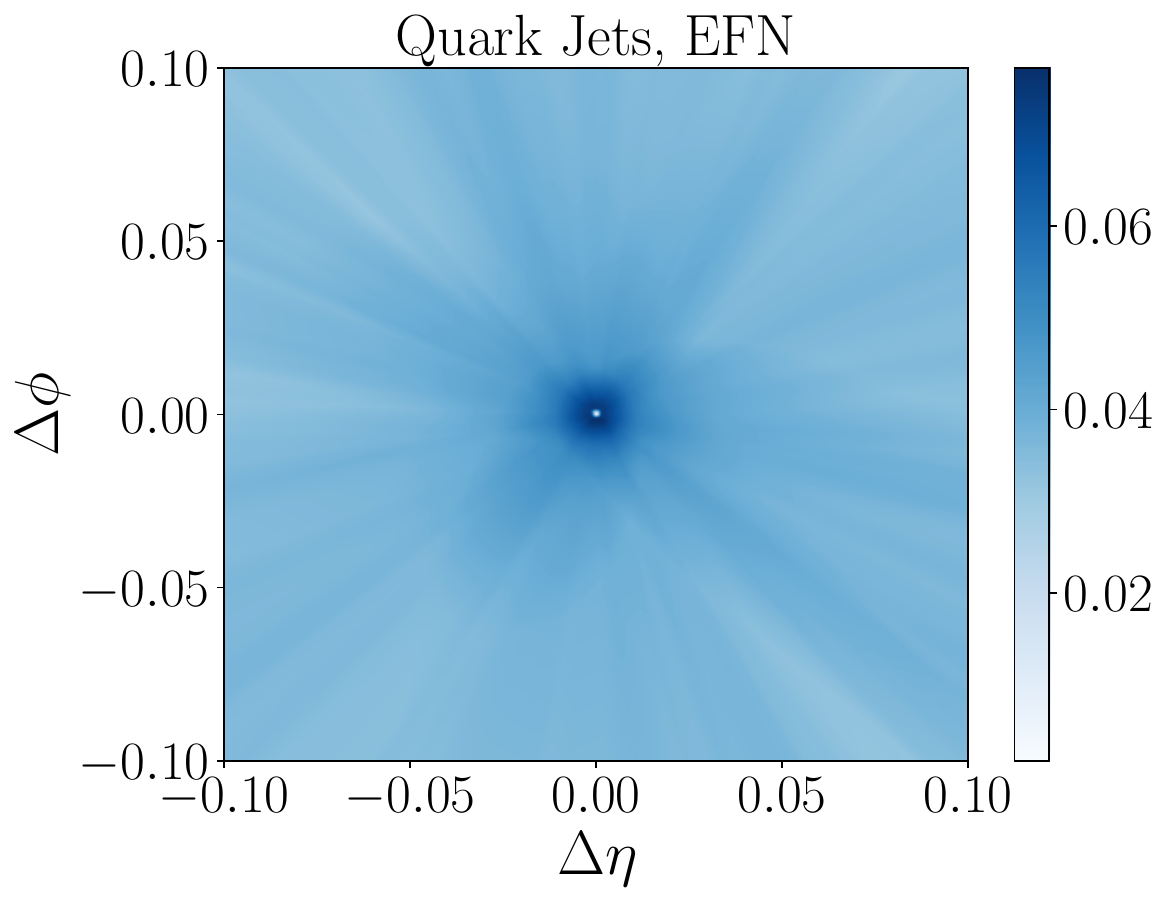}
        }
        \subfloat[][]{
        \includegraphics[width=0.5\textwidth]{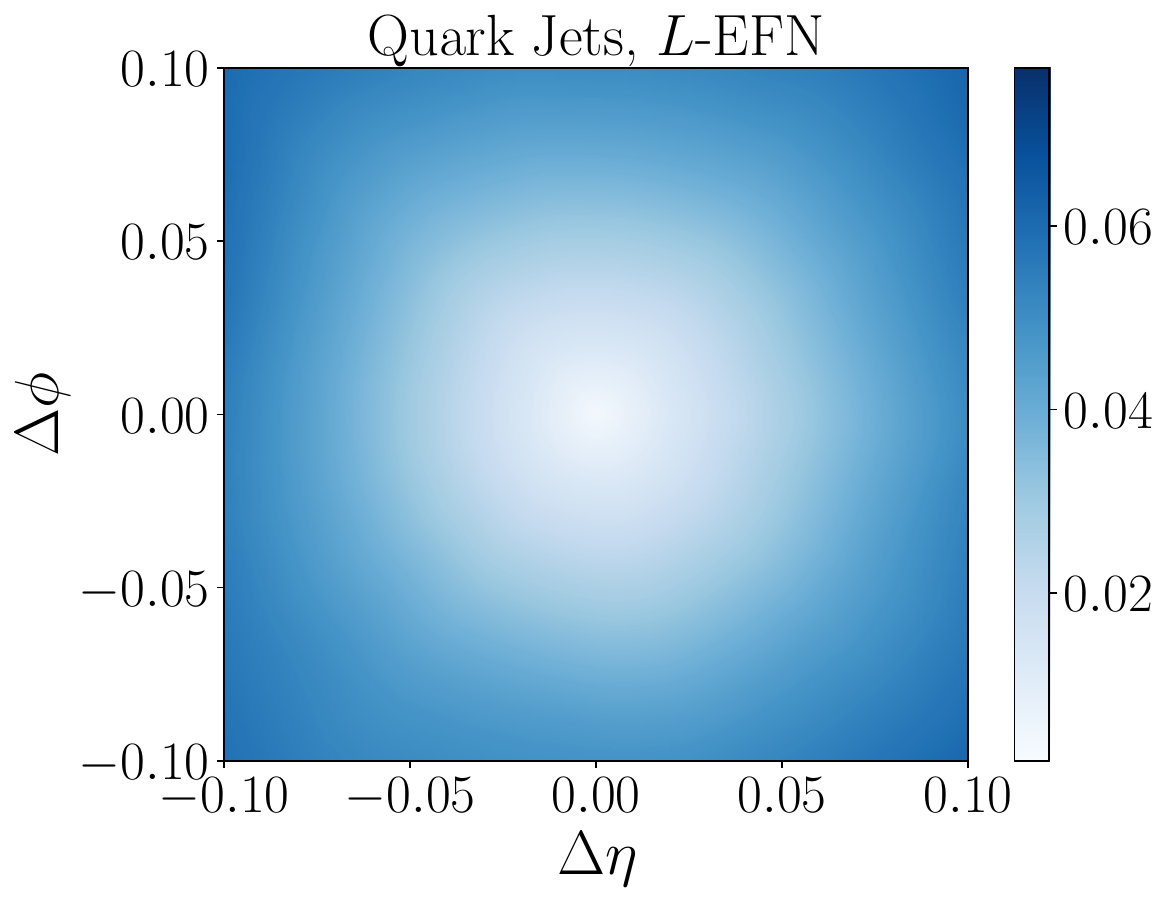}
        }
        \\
        \subfloat[][]{
        \includegraphics[width=0.5\textwidth]{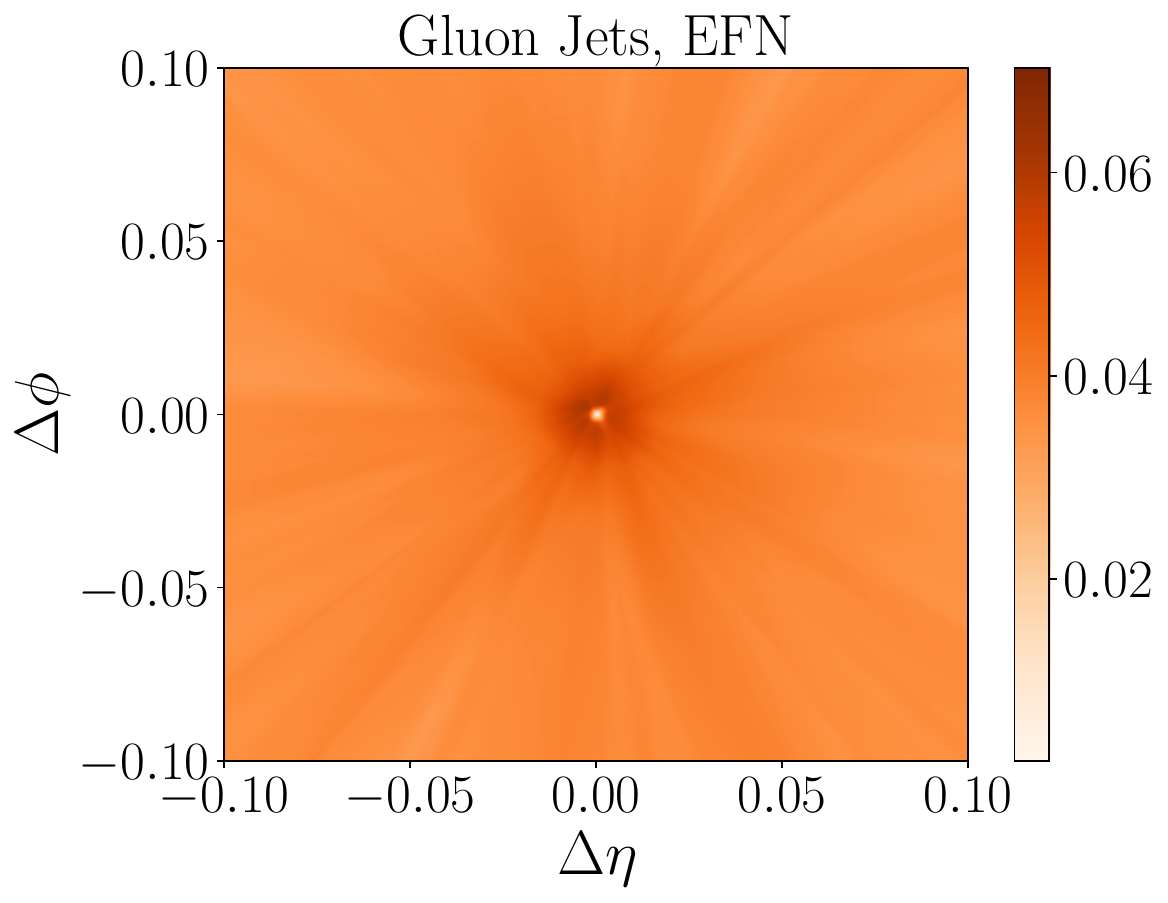}
        }
        \subfloat[][]{
        \includegraphics[width=0.5\textwidth]{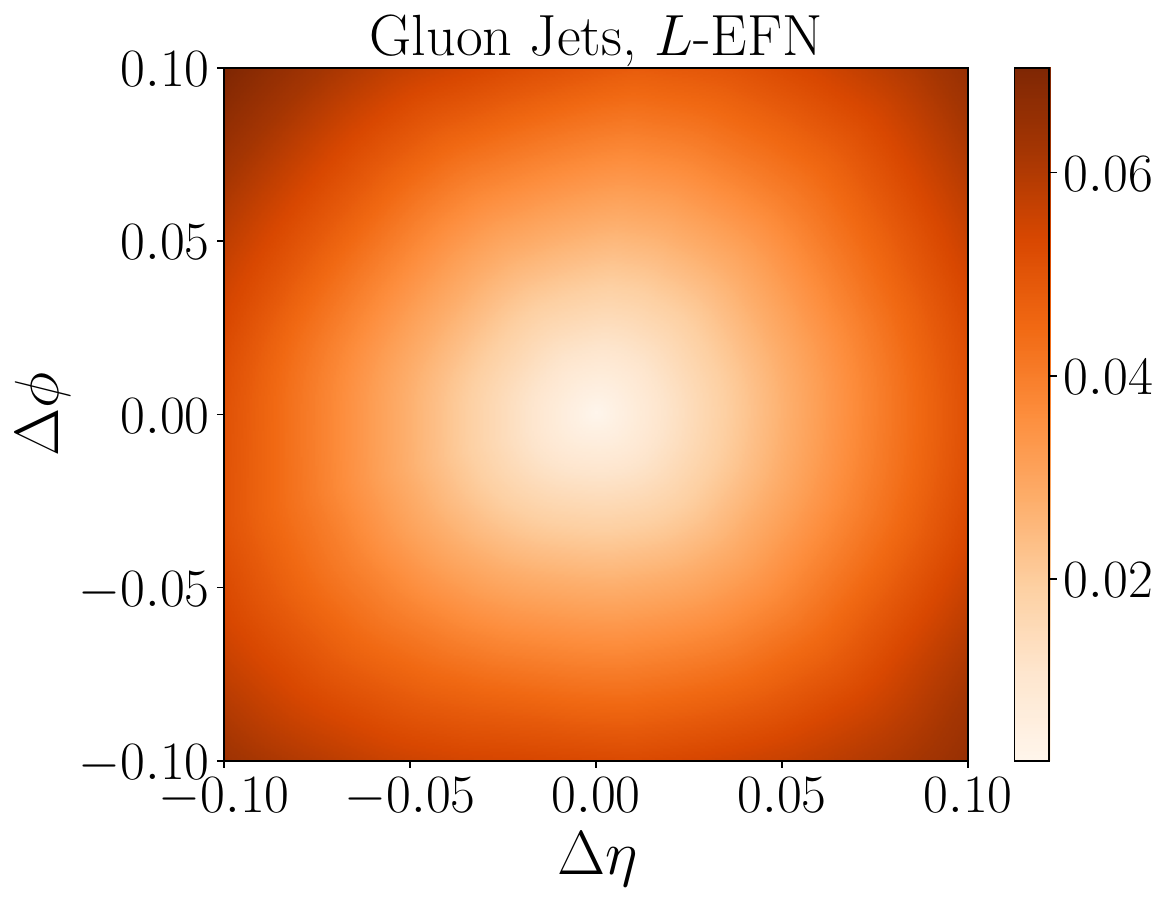}
        }
    \end{minipage}
    $\qquad$
    \begin{minipage}[c]{0.5\textwidth}
            \subfloat[][]{
    \includegraphics[width=\textwidth]{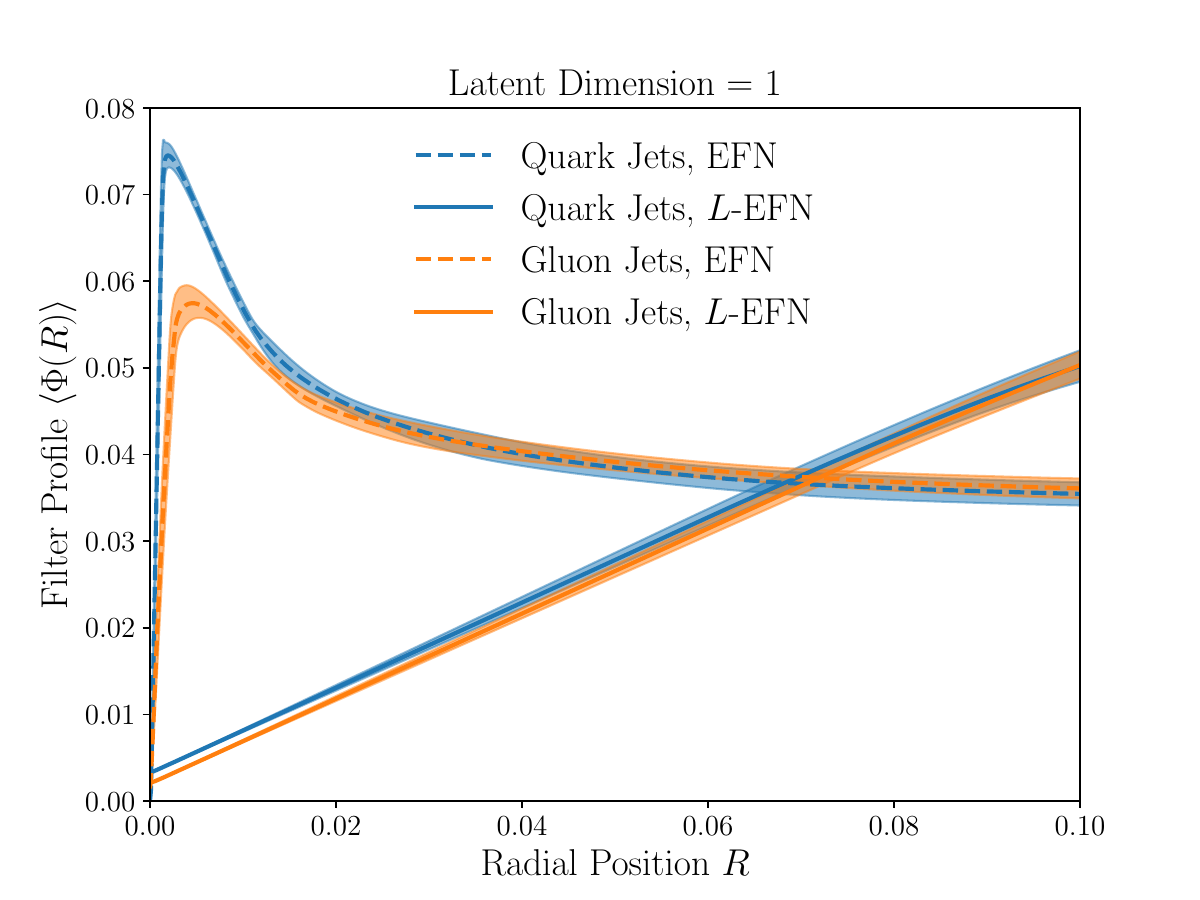}
        }
    \end{minipage}
    \caption{
    Angular filters $\Phi(\eta,\phi)$ for parton-level versus hadron-level discrimination with a latent space of dimension $\ell=1$.
    The EFN filters are shown for (a) quark and (c) gluon jets, and similarly for the $L$-EFN filters for (b) quark and (d) gluon jets.
    The radial profiles $\Phi(R = \sqrt{\eta^2 + \phi^2})$ are shown in (e), demonstrating the EFN sensitivity to highly collinear emissions ($R \lesssim 0.02$).
    While the jet radius is $R = 1.0$, we have zoomed in to show the collinear structure of the filters.}
    \label{fig:L1_filters}
\end{figure*}

\section{Visualizing Non-Perturbative Effects}
\label{sec:efn-filters}

The above case study demonstrates that IRC safety is insufficient to protect against large non-perturbative modifications, but it gives no immediate insight into which observables are most susceptible.
In this section, we take a step towards answering this question by training EFNs with a minimal latent dimension size of $\ell = 1$.
This corresponds to a single learned filter $\Phi(\eta,\phi)$ which can be easily visualized, and a simple one-dimensional classification function $F$.

As in \Sec{results}, we train $\ell=1$ EFNs and $L$-EFNs to discriminate parton- and hadron-level jets, treating the quark and gluon samples separately.
Perhaps surprisingly, the $\ell = 1$ EFNs are still able to achieve AUCs of 0.90 (quarks) and 0.91 (gluons), compared to 0.96 for the full $\ell = 60$ models in \Sec{results}.
The $\ell = 1$ $L$-EFNs perform similarly well, with AUCs of 0.54 and 0.53, compared to 0.55 and 0.54 previously.
We therefore conclude that an $\ell = 1$ analysis is sufficient to capture the leading sources of non-perturbative sensitivity.

In \Fig{L1_filters}, we plot the filter profiles $\Phi(\eta,\phi)$ for the $\ell=1$ ($L$-)EFNs (left and middle columns) alongside their angle-averaged radial profiles (right column).
All filters exhibit cylindrical symmetry, as expected since 
the generated samples are for unpolarized jets with no preferred angular orientation.
Interestingly, the EFN filters appears to be maximally sensitive to \textit{highly collinear} radiation.
Despite the jet radius being $R = 1.0$, the EFN filters peak within a radius of $\sim 0.02$, with a plateau going out to larger scales.
The EFN filters also vanish at the origin, reflecting the consistent presence of a particle at $R = 0$ aligned with the WTA jet axis.%
\footnote{If angles are instead measured with respect to the standard jet momentum axis, then the $\ell=1$ EFN filters often exhibit anisotropic features.}

The $L$-EFN filters, on the other hand, cannot vary so rapidly near the origin due to the $1$-Lipschitz constraint.
Thus, the $L$-EFN cannot be as sensitive to collinear radiation, while simultaneously ignoring the central particle. 

This differing behavior of EFNs and $L$-EFN is consistent with theoretical expectations about the non-perturbative sensitivity of the angularities:
\begin{equation}
\label{eq:angularity}
\lambda^{(\beta)} = \sum_{i\in\text{jet}}z_i\,\Delta R_{i,\text{jet}}^\beta, 
\end{equation}
which has Lipschitz constant $L = \beta$.
When $\beta \gtrsim 1$, non-perturbative corrections are suppressed by $\Lambda_{\rm QCD}/E_{\rm jet}$, so they are relatively small and independent of $\beta$ for high energy jets, in agreement with the $L$-EFN behavior.

For $\beta \lesssim 1$, though, non-perturbative corrections scale as $(\Lambda_{\rm QCD}/E_{\rm jet})^\beta$.
This is unsuppressed as $\beta \to 0$, which yields large differences between parton-level and hadron-level distributions.
Note that in the $\beta \to 0$ limit, the angularities are dominated by collinear emissions.
This explains why the IRC-safe EFN with no Lipschitz constraint can have large non-perturbative sensitivity coming from small-angle physics.

\begin{figure*}[t]
    \centering
                \subfloat[][]{
    \includegraphics[width=0.43\textwidth]{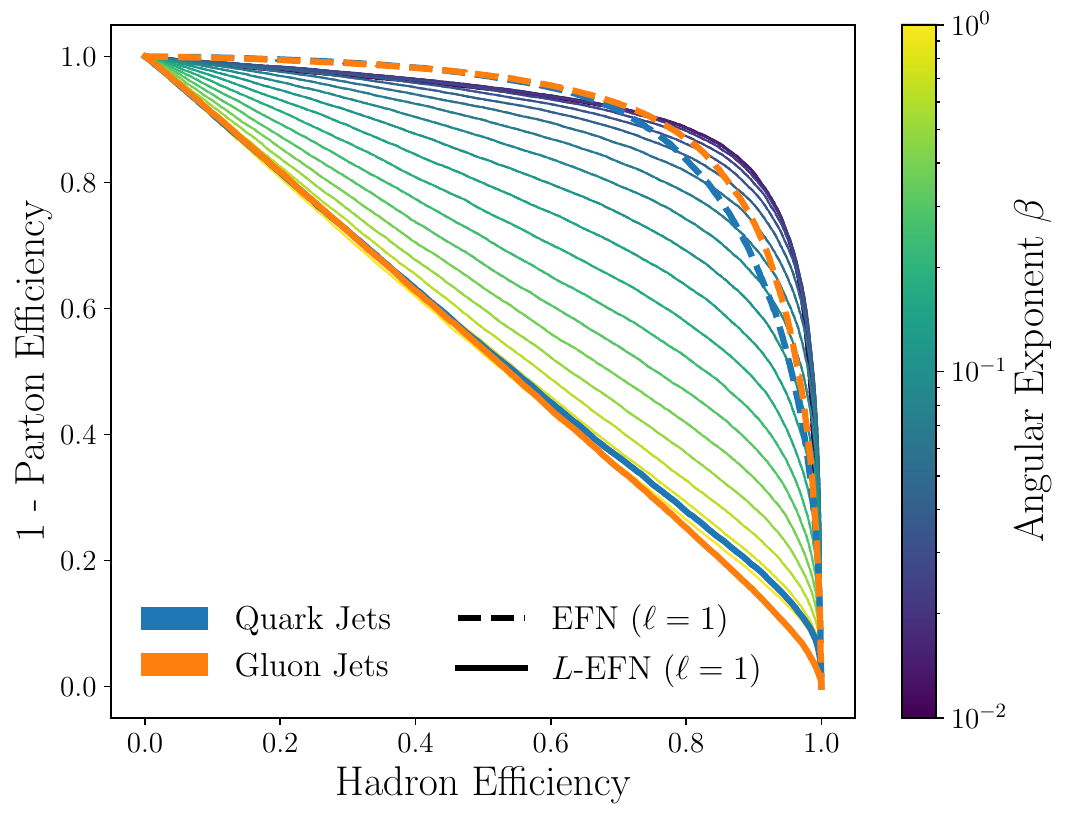}
    } $\qquad$
                \subfloat[][]{
    \includegraphics[width=0.48\textwidth]{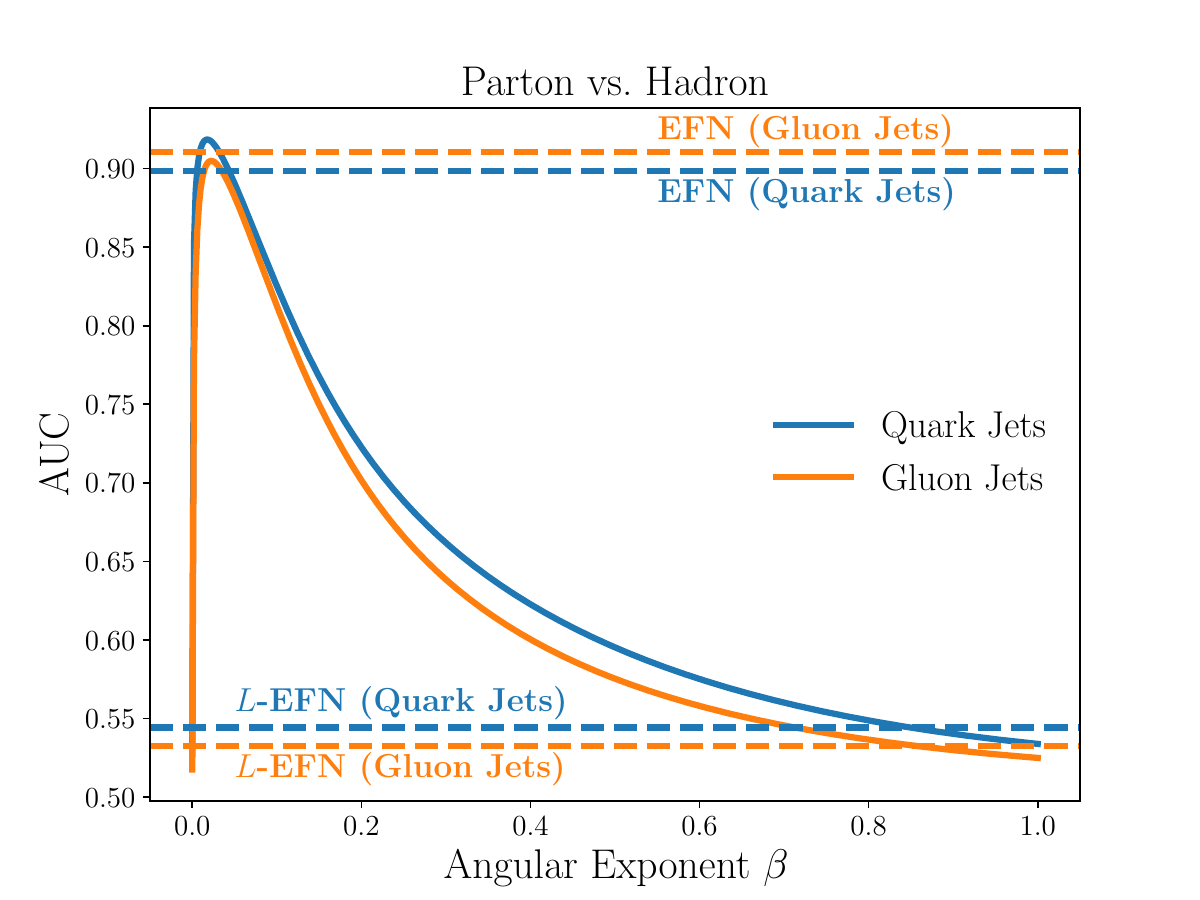}
    }
    \caption{
    Parton versus hadron discriminating power for the angularities $\lambda^{(\beta)}$ from \Eq{angularity} as a function $\beta$.
    Shown are (a) ROC curves and (b) AUC scores.
    We also show ROC curves for the $\ell = 1$ EFNs and $L$-EFNs, which agree well with the $\beta \ll 1$ and $\beta \approx 1$ limits.
    For simplicity, in (a) we only show quark ROC curves for the $\beta$ scan, though the gluon results are qualitatively similar.
    }
    \label{fig:angularities}
\end{figure*}

In \Fig{angularities}, we quantify the parton versus hadron discrimination power of the angularities as a function of $\beta$.
ROC curves are shown for the EFN and $L$-EFNs for comparison, and we see a rough interpolation where $\beta \to 0$ maps to the EFN, while $\beta \approx 1$ maps to the $L$-EFN.
The AUCs monotonically increase as $\beta$ decreases, peaking around $\beta = 0.02$ before losing sensitivity as $\beta \to 0$.

\section{Conclusions and Outlook}
\label{sec:conclusions}

In this paper, we demonstrated the susceptibility of IRC-safe neural networks to large, incalculable non-perturbative corrections.
We showed how to make the network maximally sensitive to these corrections by training it to distinguish between parton-level and hadron-level jets.
In general, there is nothing constraining this effect in a typical application.
Although IRC safety guarantees that the network's output is perturbatively calculable \textit{in principle}, this is insufficient for robustness to large and potentially mis-modeled non-perturbative corrections.

To mitigate this issue, we introduced a new neural network architecture, the $L$-EFN, which has significantly reduced sensitivity to non-perturbative physics.
We achieve this by constraining the networks to be $L$-Lipschitz functions, which bounds potential non-perturbative modifications according to the EMD between parton- and hadron-level jets.
Using \textsc{Pythia}'s hadronization model as a proxy for non-perturbative effects, we demonstrated that an $L$-EFN is barely able to distinguish between the same jet at parton- and hadron-level.
In the future, it would be interesting to explore different hadronization models and other non-perturbative effects like multiparton interactions (i.e.~underlying event).
It would also be interesting to study the robustness of $L$-EFNs to perturbative uncertainties in parton showers.

The ideas behind $L$-EFNs could be extended to other architectures, such as graph neural networks and transformers, though it may be more complicated to derive theoretical bounds similar to \Eq{efsn-bound}.
It would also be interesting to benchmark $L$-EFNs and $L$-EFN-inspired architectures on standard jet tagging tasks and assess the tradeoff between performance and robustness to hadronization.
In the case of $L$-EFNs, the Lipschitz constant provides an easily tunable network constraint that balances these considerations.
The non-Lipschitz networks trained to distinguish parton and hadron jets may also be useful to explore the impact of hadronization where it is predicted to be the largest.
For non-IRC-safe networks, our studies did not reveal dramatic differences between PFNs and $L$-PFNs, but the flexibility to adjust $L$ may be beneficial for ablation studies.

With a growing trend to build robust and interpretable neural networks that are structured with particle physics in mind, it is important to be guided by both formal and practical considerations.
Finding a balance between neat theoretical constraints and performance demands is inevitably delicate, but $L$-EFNs present a concrete step towards uniting the power of deep learning with a reasoned caution against learning spurious or unphysical details of a simulation.
We look forward to continued investigation and development in future work, and hope to see the ongoing development of new machine learning tools alongside theoretical tools to better understand them.

\begin{acknowledgments}
We thank Rikab Gambhir for feedback on the manuscript and discussions of the $\beta \to 0$ limit of angularities.
We thank Eric Metodiev and Andrew Larkoski for additional helpful comments.
SBT is supported by the National Science Foundation under Award No.\ PHY-2209443.
BN is supported by the U.S. Department of Energy (DOE), Office of Science under contract DE-AC02-05CH11231.
JT is supported by the DOE Office of High Energy Physics under Grant Contract No.\ DE-SC0012567, by the National Science Foundation under Cooperative Agreement No.\ PHY-2019786 (The NSF AI Institute for Artificial Intelligence and Fundamental Interactions, \url{http://iaifi.org/}), and by the Simons Foundation through Investigator grant 929241.
This research used resources of the National Energy Research Scientific Computing Center, a DOE Office of Science User Facility supported by the Office of Science of the U.S. Department of Energy under Contract No. DE-AC02-05CH11231 using NERSC award HEP-ERCAP0021099.
\end{acknowledgments}

\appendix

\section{Quark vs.\ Gluon Jet Discrimination with $L$-EFNs}
\label{app:quarkgluon}

In \Secs{casestudy}{efn-filters}, we contrasted EFNs and $L$-EFNs for the relatively contrived scenario of distinguishing parton- from hadron-level jets.
While this demonstrated our point about the large potential impact of non-perturbative physics on EFN observables, it does not reflect a realistic use case of ($L$-)EFNs.
In this appendix, we compare EFNs and $L$-EFNs on a canonical collider physics task: discriminating quark and gluon jets.

\begin{figure*}[t]
    \centering
            \subfloat[][]{
    \includegraphics[width=0.45\textwidth]{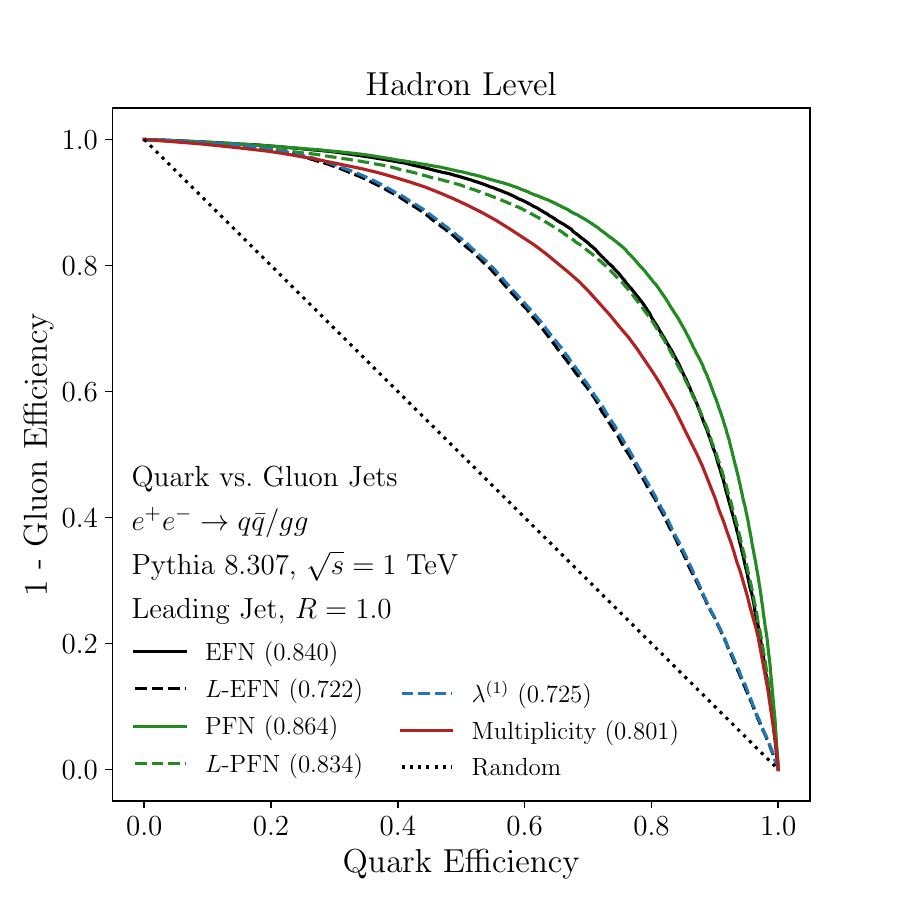}
    } $\qquad$
            \subfloat[][]{
    \includegraphics[width=0.45\textwidth]{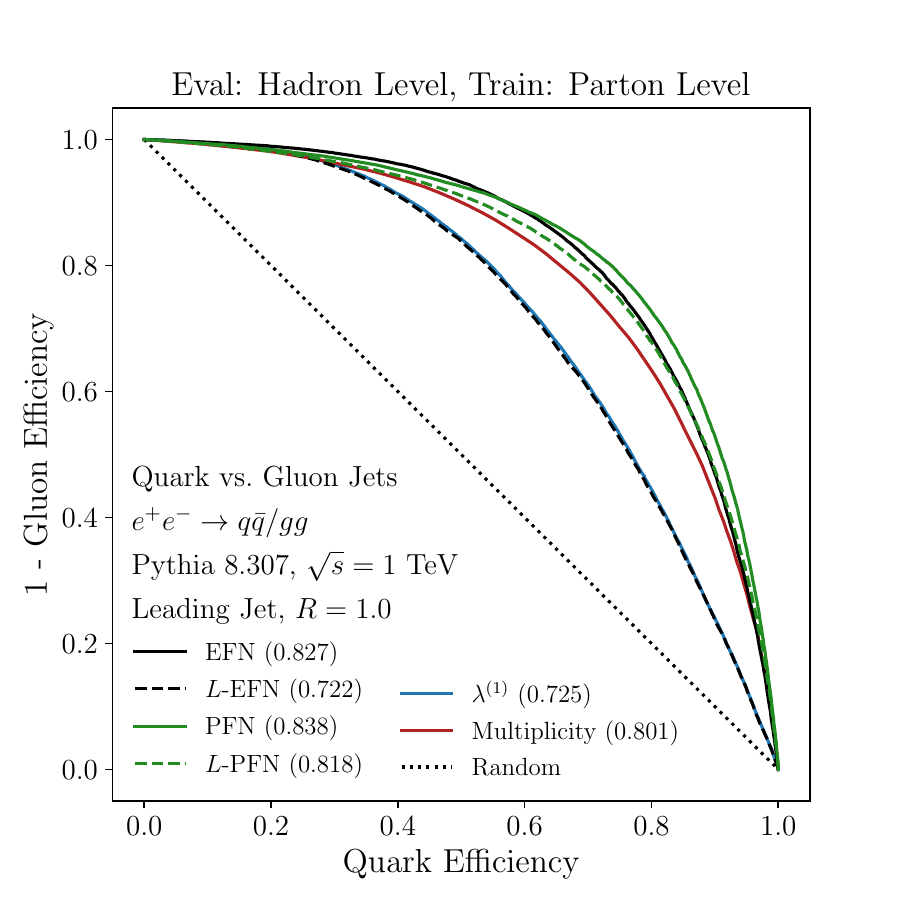}
    }
    \caption{
    ROC curves for hadron-level quark/gluon jet discrimination using an EFN/PFN (solid) and an $L$-EFN/$L$-PFN (dashed), trained using (a) hadron-level and (b) parton-level inputs.
    The results for constituent multiplicity and the angularity $\lambda^{(1)}$ are shown for reference.
    While the $L$-EFN underperforms with respect to the EFN, it is minimally affected by training at parton-level and evaluating at hadron-level.
    The EFN loses a small amount of discriminating power, and the PFN/$L$-PFN lose even more.
    }
    \label{fig:qg_EFN:roc}
\end{figure*}

Using the datasets described in \Sec{dataset}, we train different ($L$-)EFNs and ($L$-)PFNs to discriminate quark jets from gluon jets.
We perform separate trainings on hadron-level and parton-level inputs, but always evaluate the performance on hadron-level jets.
In \Fig{qg_EFN:roc}, we show ROC curves for quark/gluon jet discrimination using these ($L$)-EFNs/PFNs.
As expected, the default $L$-EFN with $L = 1$ performs exactly as well as the angularity $\lambda^{(1)}$, both of which underperform relative to the unconstrained EFN and both PFNs.
Notably, the $L$-EFN performance is nearly identical whether it is trained on hadron- or parton-level inputs, highlighting the expected robustness to nonperturbative effects.

\bibliography{refs}

\end{document}